\documentclass[]{aa} 
\usepackage{graphicx}
\usepackage{natbib}
\usepackage{txfonts}
\begin{document}
   \title{Oscillation mode lifetimes in $\xi\,$Hydrae: Will strong mode damping limit asteroseismology of red giant stars?}

\titlerunning{Will strong mode damping limit asteroseismology of red giant stars?}

   \author{D. Stello\inst{1,2,3}
          \and
          H. Kjeldsen\inst{1}
          \and
          T. R. Bedding\inst{2}
          \and
          D. Buzasi\inst{3}
          }

   \offprints{D. Stello: stello@phys.au.dk}

   \institute{Institute for Fysik og Astronomi (IFA), Aarhus Universitet,
              8000 Aarhus, Denmark
         \and
             School of Physics, University of Sydney,
             NSW 2006, Australia  
         \and
             Department of Physics, US Air Force Academy,
             Colorado Springs, CO 80840, USA
             }


   \date{Received 10 August 2005 / accepted 4 November 2005}

   \abstract{We introduce a new method to measure frequency separations and
             mode lifetimes of stochastically excited and
             damped oscillations, so-called solar-like oscillations.
             Our method shows that velocity data of the red giant star
             $\xi\,$Hya \citep{Frandsen02} support
             a large frequency separation between modes 
             of roughly $7\,\mu$Hz.
             We also conclude that the data are consistent with a mode
             lifetime of 2 days, which is so short relative to its
             pulsation period that none of the observed frequencies are
             unambiguous.
             Hence, we argue that the maximum asteroseismic
             output that can be obtained from these data is an average large
             frequency separation, the oscillation amplitude and the average 
             mode lifetime.
             However, the significant discrepancy between the theoretical
             calculations of the mode lifetime \citep{HoudekGough02} and
             our result based on the observations of
             $\xi\,$Hya, implies that red giant stars can help us
             better understand the damping and driving mechanisms of 
             solar-like p-modes by convection.

   }

   \maketitle
%

\section{Introduction}

The mode lifetime of solar-like oscillations is an important
parameter.
The interpretation of the measured oscillation frequencies (and
their scatter) relies very much on knowing the mode
lifetime, but currently we know very little
about how this property depends on
the stellar parameters (mass, age and chemical composition).
The theoretical estimates of mode lifetimes are based on
a simplified description of the convective environment in which
the damping and excitation of the modes takes place.
Measurements of the mode lifetime, $\tau$, in different stars will be very
helpful for a more
thorough treatment of convection in stellar modeling.
In this paper the mode lifetime refers to the time for 
the amplitude to decrease by a factor of $e$.

The number
of measurements of the mode lifetime, or damping time, is still very limited.
Observations of main-sequence stars imply mode lifetimes of a few days in
the Sun \citep{Libbrecht88,Chaplin97}, $\alpha\,$Cen A \citep{Bedding04}
and $\alpha\,$Cen B \citep{Kjeldsen05}.
As pointed out by \citet{Bedding05}, independent observational
studies on the star Procyon
do not agree on the measured frequencies, a disagreement that could be
the result of a short mode lifetime. The power spectrum of
the K giant Arcturus reported by \citet{Retter03} could also be
explained as a short mode lifetime ($\tau=2\,$days) of a single mode.
If the mode lifetime does not increase with oscillation period,
this would limit the prospects of asteroseismology on
evolved stars that have periods of several hours or
longer, because of poor coherence of the oscillations.
Only when we look at M giants -- the semi-regular variables -- do we
see evidence for longer lifetimes, ranging from years to decades
\citep{Dalsgaard01,Bedding03,Bedding05a}.

The theoretical predictions of mode lifetimes for unevolved
stars like the Sun \citep{Houdek99} and $\alpha$ Cen A \citep{Samadi04}
are in a fairly good agreement with the observed values.
However, for the more evolved red giant star $\xi\,$Hya, there
seems to be a significant discrepancy between theory 
\citep[$\tau\sim15$--20$\,$days;][]{HoudekGough02} and observation 
\citep[$\tau\sim2$--3$\,$days;][ hereafter Paper I]{Stello04}.
In Paper I we also measured the oscillation
amplitude to be roughly $2\,$ms$^{-1}$, which was in good agreement with
the theoretical value \citep{HoudekGough02}.

\begin{figure}\centering
\includegraphics{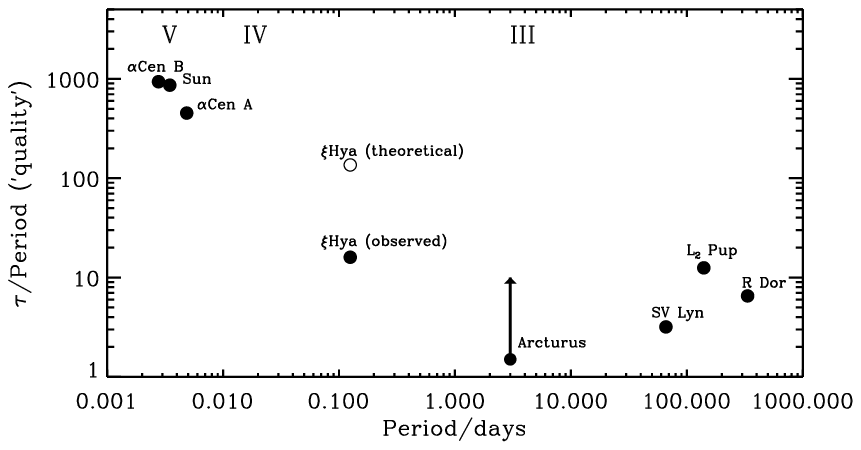}
\caption[quality]{\label{figteas}
{The oscillation `quality' factor vs. period for selected stars. Filled 
symbols are measured, while the empty symbol shows the theoretical value.
The arrow indicates a lower limit. Luminosity classes are indicated at 
the top.
The measured values are from: $\alpha\,$Cen B, \citet{Kjeldsen05}; 
Sun, \citet{Chaplin97}; $\alpha\,$Cen A, \citet{Kjeldsen05}; $\xi\,$Hya 
(theoretical), \citet{HoudekGough02}; $\xi\,$Hya 
(observed), Paper I; Arcturus, \citet{Retter03}; 
L$_2$ Pup, \citet{Bedding05a}; SV Lyn and R Dor, \citet{Dind04}.}}
\end{figure}
In Fig.~\ref{figteas} we plot the measured ratios between the mode 
lifetime and period (the oscillation `quality' factor) as a function 
of period for selected stars, including the theoretical value for
$\xi\,$Hya. Roughly speaking, the `quality' factor is the number
of oscillation cycles with constant phase, and the higher this number, 
the better we can determine the frequency. 
Note that the relation between the mode lifetime 
$\tau$ and the the FWHM $\Gamma$ (in cyclic 
frequency) of the corresponding resonant 
peak is $\Gamma=1/(\pi\tau)$.

In this paper we further investigate the mode lifetime of $\xi\,$Hya
and examine whether it limits the
possible astrophysical output.
We introduce a new method that measures repeated
frequency patterns (e.g. the large frequency separation,
$\Delta\nu_{0}$) and the mode lifetime of solar-like p-mode
oscillations.
We furthermore analyze the stellar power spectrum to establish
the most likely mode lifetime, by comparing
the cumulative power distribution of the observations with
simulations. Finally, the ambiguity of the measured frequencies
is quantified.


\section{The $\xi\,$Hya data set and previous results}
\begin{figure}\centering
\includegraphics{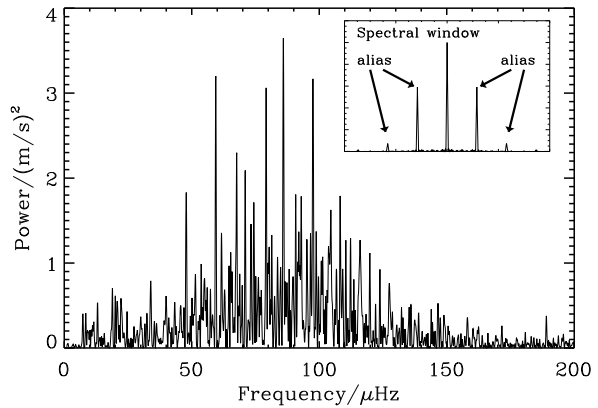}
\caption[scaled solar spectrum]{\label{fig4}
{Power spectrum of $\xi\,$Hya. The frequency axis of the
spectral window in the inset is scaled to match that of the main plot.
}}
\end{figure}

We use the same data set as in Paper I, which
comprises 433 measurements of radial velocity
covering almost 30 days of single-site observations using the
\textsc{Coralie} spectrograph at La Silla (ESO, Chile).
The average noise per measurement is
$\sigma_{\mathrm{measure}}=2.33\,\mathrm{ms}^{-1}$, and the data provide a
clear detection of excess power in the Fourier spectrum at roughly
90$\,\mu$Hz (period $\sim3\,$hours) with many peaks of S/N$>$3
(see Fig.~\ref{fig4}). For further details about the observations
see \citet{Frandsen02}.
Note that during the reduction the data was high-pass filtered 
\citep{Frandsen02}, but we have shown with simulations that this does 
not affect the power spectrum above $20\,\mu$Hz.
Using the autocorrelation of the power spectrum
\citet{Frandsen02} found a
frequency spacing of $6.8\,\mu$Hz, which is in good agreement with
the large frequency separation, $\Delta\nu_{0}$, from a pulsation model
of the star \citep{Dalsgaard04}.
For further analysis on the stellar parameters of $\xi\,$Hya see e.g.
\citet{Stello02,Frandsen02,Teixeira03}; Paper I; \citet{Thevenin05}.

In Table~\ref{table1} we give the frequencies used in the current
investigation. We used the conventional method of
iterative sine-wave fitting (`prewhitening') 
to measure the 10 frequencies listed in Table~\ref{table1} 
(see Appendix~\ref{clean}).
The uncertainties
are calculated according to \citet{MontgomeryODonoghue99}
\citep[see also][]{Kjeldsen03}.
\begin{table}
\centering{{\footnotesize
\caption[Extracted peaks from $\xi\,$Hya power spectrum]{Measured frequencies and amplitudes for $\xi\,$Hya}
\label{table1}
 \begin{tabular}{lr@{.}lr@{.}lr@{.}lr@{.}l}
 \hline
 \hline
 ID & \multicolumn{2}{c}{Frequency} &
 \multicolumn{2}{l}{Amplitude}  & \multicolumn{2}{l}{S/N} & \multicolumn{2}{l}{Frandsen et al.}\\
    & \multicolumn{2}{c}{$\mu$Hz} &
 \multicolumn{2}{c}{ms$^{-1}$}  & \multicolumn{2}{l}{} & \multicolumn{2}{c}{2002}\\
 \hline
 \noalign{\smallskip}
  $\nu_1$     &  85&91(3)$^{a}$  & 1&89(24)$^{a}$ & 6&3 &  85&96(3)\\
  $\nu_2$     &  59&43(3)        & 1&75           & 5&8 &  59&43(3)\\
  $\nu_3$     &  79&13(3)        & 1&65           & 5&5 &  79&13(3)\\
  $\nu_4$     &  95&21(4)        & 1&33           & 4&4 &  95&28(4)\\
  $\nu_5$     & 108&20(4)        & 1&24           & 4&1 & 108&22(4)\\
  $\nu_6$     & 101&32(4)        & 1&17           & 3&9 & 101&16(4)\\
  $\nu_7$     &  98&76(4)        & 1&12           & 3&7 &  98&77(5)\\
  $\nu_8$     & 112&29(5)        & 1&11           & 3&7 &  --&--   \\
  $\nu_9$     & 105&14(5)        & 1&01           & 3&4 & 105&13(5)\\
  $\nu_{10}$  &  73&38(5)        & 1&02           & 3&4 &  --&--   \\
 \hline
  \multicolumn{7}{l}{\raisebox{1ex}{\tiny{$a$}} \scriptsize{The errors are
   derived from: $\sigma_{f}=\sqrt{6}/(T_{\mathrm{obs}}\cdot$S/N$\cdot\pi^{3/2})$,}}\\
  \multicolumn{7}{l}{ \scriptsize{and $\sigma_{\mathrm{amp}} =
 \sqrt{2/\pi}\langle\mu_{\mathrm{amp}}\rangle$, where
 $\langle\mu_{\mathrm{amp}}\rangle=\sigma_{\mathrm{measure}}\sqrt{\pi/N_{\mathrm{obs}}}$,}} \\
  \multicolumn{7}{l}{ \scriptsize{and $N_{\mathrm{obs}}=433$ (see text).}} \\
 \end{tabular}
}}
\end{table}
We note that these frequencies are not exactly the same as those
quoted by \citet{Frandsen02}, who used a different method to
measure the frequencies. This discrepancy is not
surprising because the alias peaks (Fig.~\ref{fig4}) are likely to 
introduce small frequency
differences, depending on the method used to extract the frequencies.
The uncertainties indicated in 
Table~\ref{table1} are based on the signal-to-noise (S/N) and
do not take into account the scatter of the frequencies 
arising from a short mode lifetime.
This extra frequency scatter is the main subject of this paper.


\section{The method}\label{method}


\subsection{Background and definitions}\label{def}
From theoretical calculations we expect the oscillations in red giant stars
to be dominated by radial modes
\citep{Dalsgaard04,Guenther00,Dziembowski01}, which is also supported by
the observations of $\xi\,$Hya (Paper I) and other red giants
\citep{Buzasi00,Kiss03,Retter03}.
We therefore assume that the frequencies,
$\nu_n$, of $\xi\,$Hya will show a simple comb pattern in the power
spectrum with only radial modes.
This can be described by the linear relation
\begin{equation}
  \nu_{n}\simeq \Delta\nu_{0}n+X_{0}\,,
\label{eq_asymtot1}
\end{equation}
where $\Delta\nu_{0}$ is the separation between mode frequencies of
successive order $n$ and $X_{0}$ is an offset.

The finite lifetimes of the oscillation modes will introduce deviations
of the measured frequencies from the true mode
frequencies and hence from the regular comb pattern
\citep{Anderson90,Bedding04}.
\begin{figure}\centering
\includegraphics{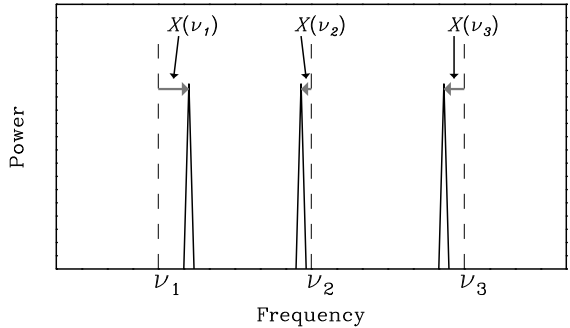}
\caption[quality]{\label{figoffset}
{Schematic illustration of the frequency scatter due to a finite 
mode lifetime. Dashed lines are the true mode frequencies 
(Eq.~\ref{eq_asymtot1}), and 
solid peaks are measured frequencies.}}
\end{figure}
In Fig.~\ref{figoffset} we illustrate this by showing, schematically, 
the comb pattern of `true' mode frequencies (dashed lines; 
Eq.~\ref{eq_asymtot1}) and the 
measured frequencies (solid peaks). The deviations, indicated with
$X(\nu_{i})$ (Eq.~\ref{mod_func}), are independent because each mode 
is excited independently.
Shorter mode lifetimes will give larger deviations.
Following \citet{Bedding04} and \citet{Kjeldsen05},
we estimate the mode lifetime from the scatter of the measured
frequencies about a regular pattern.
Our method differs by using a comb pattern for the reference
frequencies, and by not requiring us to assign the mode order or degree
to the measured frequencies. This is an advantage when the power
spectrum is crowded due to aliasing and noise, as in this case,
but the drawback is lower sensitivity to the mode lifetime.

The first step is to find the comb pattern that best matches our
measured frequencies,
$\nu\equiv[\nu_{1},\dots,\nu_{\mathrm{N}}]$. We do this by minimizing
the RMS difference between the measured frequencies and those given by
Eq.~\ref{eq_asymtot1}, with $\Delta\nu_{0}$ and $X_{0}$ as free parameters.
The individual deviations are
\begin{equation}
X(\nu_{i})\equiv (\nu_{i}-X_{0})\bmod \Delta\nu_{0}\,,
\label{mod_func}
\end{equation}
where we use a modulo operator that returns values between
$-\frac{1}{2}\Delta\nu_{0}$ and $\frac{1}{2}\Delta\nu_{0}$.
The RMS scatter of $X$ is then
\begin{equation}
\sigma_{X}(\nu;\Delta\nu_{0},X_{0})
                = \frac{\sqrt{2}}{\Delta\nu_{0}}
                  \sqrt{\frac{1}{N} \sum_{i=1}^{N} X(\nu_{i})^{2}
                       }\,.
\label{mod_func_stdev}
\end{equation}
Here, $N=10$ is the number of frequencies.
The normalization factor,$\sqrt{2}/\Delta\nu_{0}$, is included to
make $\sigma_{X}$ independent of $\Delta\nu_{0}$ for
randomly distributed frequencies.
We find the minimum of $\sigma_{X}$, min($\sigma_{X}$), using the Amoeba
minimization method \citep{Press92}, and calibrate it against
simulations with known mode lifetime, as described in the next section.
As a by-product, we also get estimates for $\Delta\nu_{0}$ and $X_{0}$.


\subsection{Simulations and calibration}\label{sim}

\begin{figure}\centering
\includegraphics{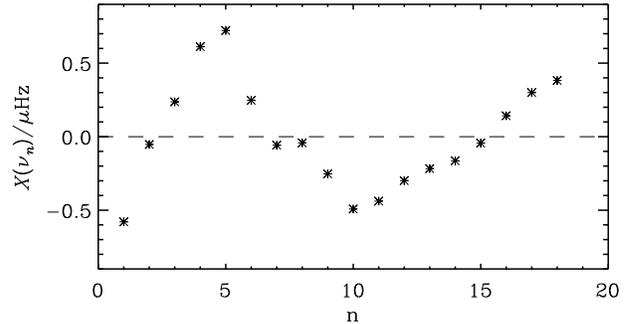}
\caption[Input frequencies]{\label{fig0}
{Deviation of input frequencies from a comb pattern (Eq.~\ref{mod_func}).
}}
\end{figure}

We simulated the $\xi\,$Hya time series using the method
described in Paper I. The oscillation mode lifetime was an
adjustable parameter, assumed to be independent of frequency,
while the other inputs for the simulator were fixed and
chosen to reproduce the observations (see Paper I, Fig.~12).
To make the simulations as realistic as possible,
the input frequencies were the radial modes from a pulsation
model of $\xi\,$Hya \citep{Teixeira03,Dalsgaard04}.
Since our model assumes the frequencies to
be strictly regular (Eq. \ref{mod_func_stdev}), the intrinsic
deviation of the input frequencies from a comb pattern
(see Fig.~\ref{fig0}) will contribute
to $\sigma_{X}$. However, as described in Sect.~\ref{robust},
this contribution turns out to be negligible and therefore
does not affect our measurement of the mode lifetime in $\xi\,$Hya.

\begin{figure}\centering
\includegraphics{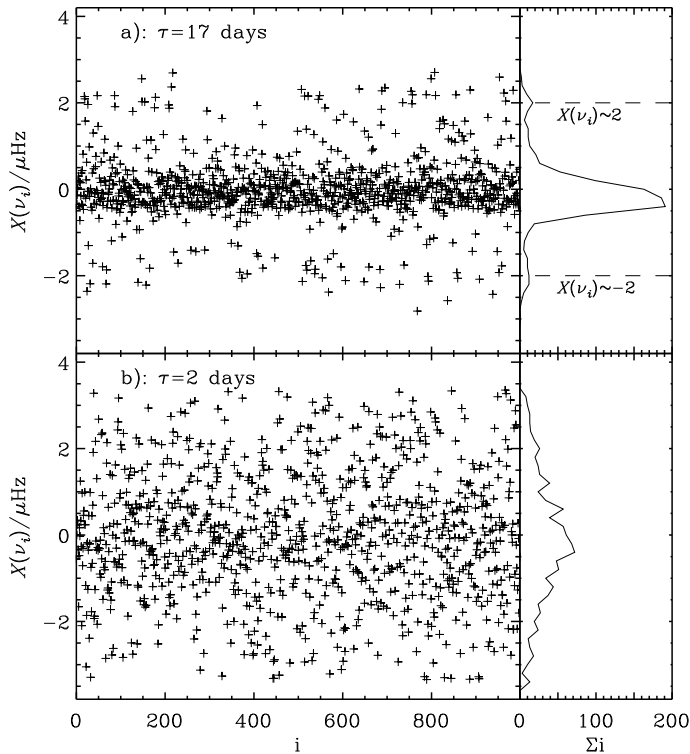}
\caption[modulo-diagram]{\label{fig1}
{The deviation from a comb pattern (Eq.~\ref{mod_func})
for 10 measured frequencies from each of 100 simulated time series.
The two horizontal bands (at $X(\nu_{i})\sim2\,\mu$Hz
and $-2\,\mu$Hz, most evident in panel (a))
below and above the central band are signatures of the
alias peaks that are present in our spectral window
(see Fig.~\ref{fig4}). To the right, the
distributions are shown for each panel.
}}
\end{figure}

\begin{figure}\centering
\includegraphics[height=15cm]{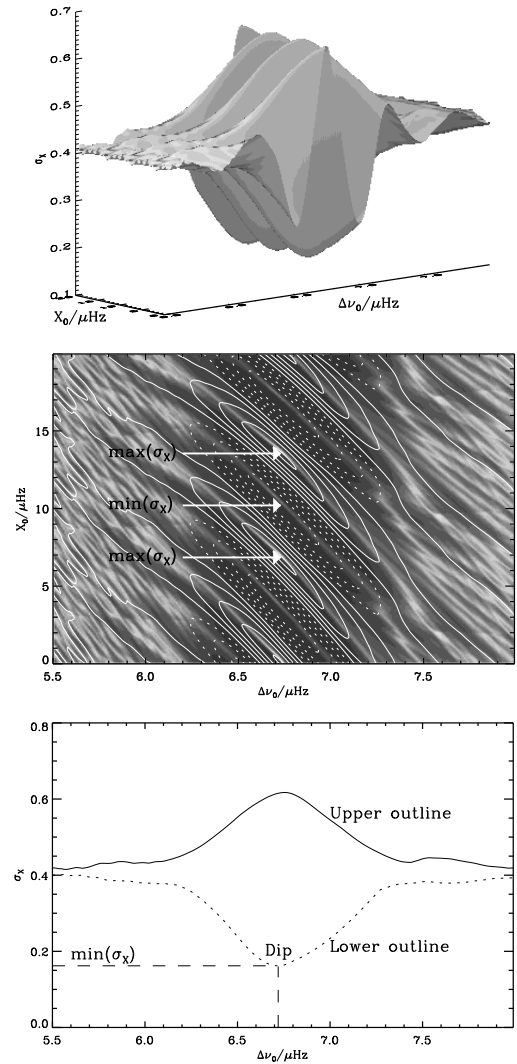}
\caption[3Dsurface]{\label{fig2}
{\textbf{Top panel:} 3D surface of $\sigma_{X}$,
(averaged from 100 simulations with $\tau=17\,$days) as a function of the
free parameters $X_{0}$ (plot range 0,20) and $\Delta\nu_{0}$
(plot range 5.5,8.0).
\textbf{Middle panel:} The surface of $\sigma_{X}$ viewed from above, where
contours are plotted to guide the eye. Solid lines indicate
contours above the mean level, while the dotted lines are those below.
\textbf{Bottom panel:} The outline of the 3D surface viewed edge on in the
$X_{0}$-direction. The two output parameters of the minimization
process, min($\sigma_{X}$) and the corresponding
$\Delta\nu_{0}$, are indicated by dashed lines.
}}
\end{figure}

\begin{figure}\centering
\includegraphics{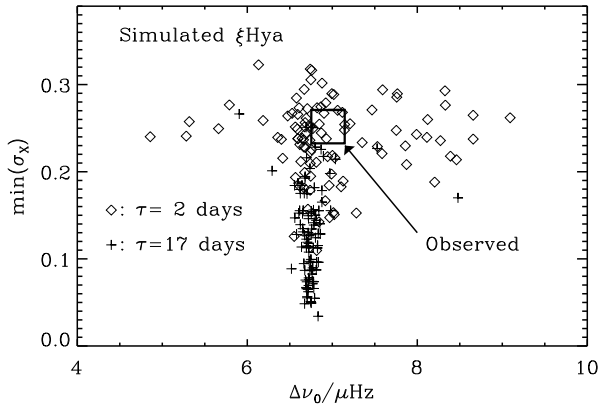}
\caption[sim data]{\label{fig5}
Output from our method applied on 100 simulations of
$\xi\,$Hya with mode lifetime 2 days (diamonds), and 17 days
(plus symbols). The square shows the observational point, and the
size of the symbol indicates the uncertainty according to the
chosen CLEAN algorithm (cf. Sect.~\ref{robust}).
}
\end{figure}

\begin{figure}\centering
\includegraphics{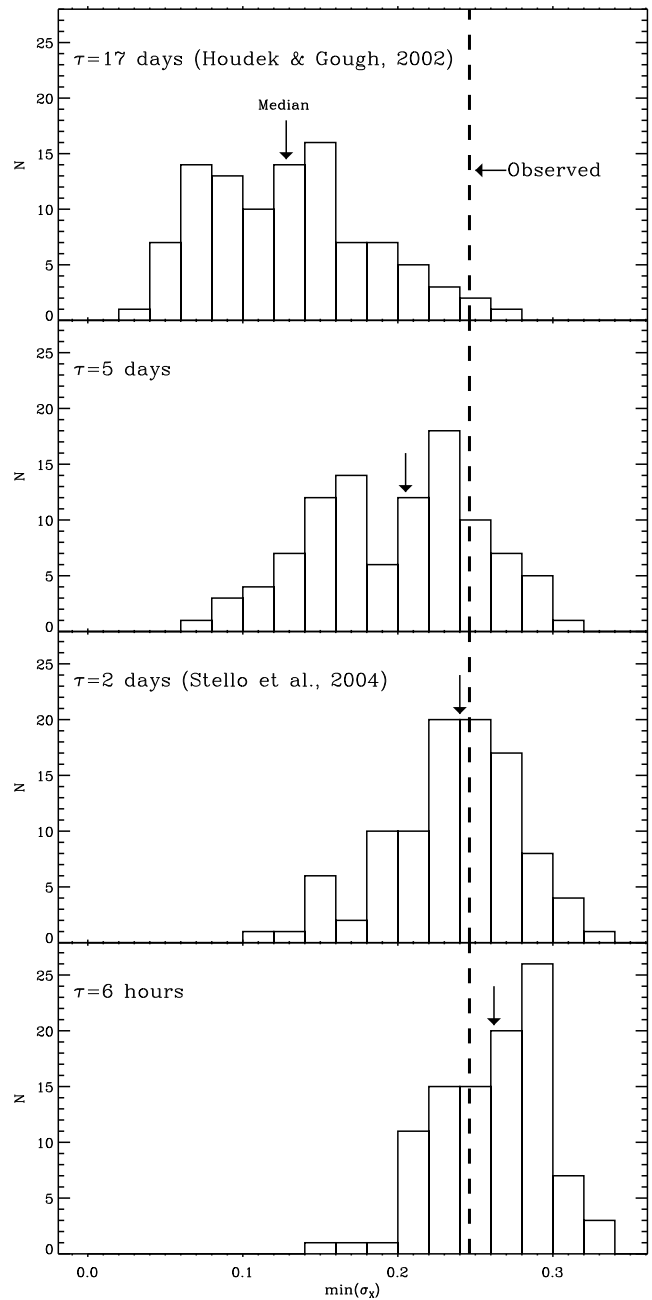}
\caption[S(nu) distrib]{\label{fig3}
{Distribution of the minimized scatter, $\sigma_{X}$,
based on 10 frequencies from each of 100 simulations with different
mode lifetimes: 17 days, 5 days, 2 days, and 6 hours.
The dashed line indicates the value from the observations, and
the vertical arrows show the median.
}}
\end{figure}

For different values of the mode lifetime, we simulated 100 time series
with different random number seeds. For each we
measured 10 frequencies ($\nu_{1},\dots,\nu_{\mathrm{10}}$) using
iterative sine-wave fitting (see Appendix~\ref{clean})
and then minimized $\sigma_{X}$.
This provided 100 values of
min($\sigma_{X})$ for each mode lifetime, which
can be compared with the observations.
In Figs.~\ref{fig1} and \ref{fig2} we illustrate
the method, and Figs.~\ref{fig5} and \ref{fig3} show the results.
Fig.~\ref{fig1} shows $X(\nu_{i})$ of all 1000 frequencies
(10 frequencies from each of 100 simulated time series) for two
mode lifetimes, (a) $\tau=17$ days, and (b) $\tau=2$ days.
The difference in mode lifetime is clearly reflected in the difference
in the scatter of $X(\nu_{i})$.
We note that the distribution of $X(\nu_{i})$ (right panels) is
affected by the presence of false detections of
alias peaks from neighbouring modes at roughly $X(\nu_{i})\sim2\,\mu$Hz
and $-2\,\mu$Hz (see Fig.~\ref{fig1}a).

In Fig.~\ref{fig2} we show $\sigma_{X}(\nu;\Delta\nu_{0},X_{0})$
for the case of $\tau=17\,$days.
To smooth the plotted surface we show the average of all 100
simulations.
However, during the minimization of $\sigma_{X}$, each
simulations was treated separately.
Note that $\sigma_{X}(\nu;\Delta\nu_{0},X_{0})=
\sigma_{X}(\nu;\Delta\nu_{0},X_{0}+n\Delta\nu_{0})$, where $n$
is an integer, and that $\sigma_{X}$ has
maxima and minima for roughly the same $\Delta\nu_{0}$
that are separated by $\frac{1}{2}\Delta\nu_{0}$ on the
$X_{0}$-axis.

The output parameters min($\sigma_{X}$) and $\Delta\nu_{0}$ are
plotted in Fig.~\ref{fig5}.
We see clearly that a smaller min($\sigma_{X}$) (more prominent
comb pattern) gives a more accurate $\Delta\nu_{0}$ determination.
In the case of short mode lifetime, the frequency
pattern is generally not very pronounced (high min($\sigma_{X}$))
mainly due to the false detections from alias peaks.

Due to the stochastic variations in the simulated time series,
the value of min($\sigma_{X}$) (Fig.~\ref{fig5}) shows a large 
intrinsic scatter from one simulation
to the next. This tells us the precision with which we can
determine the mode lifetime, $\tau$, from a single data set.
For each $\tau$, we compared the observed
min($\sigma_{X}$) with the
distribution from 100 independent simulations. This gives a measure of
whether the observed value is consistent with $\tau$.

Fig.~\ref{fig3} shows the distributions of
min($\sigma_{X}$) for simulations
as histogram-plots, for different mode lifetimes
($\tau=17\,$days, $5\,$days, $2\,$days, and $6\,$hours).
The dashed line indicates the observed value.
For a mode lifetime of 17 days (corresponding to the
damping rate in cyclic frequency, plotted by \citealt{HoudekGough02},
of $0.1\,\mu$Hz) only a few out of 100 trials
have a value for min($\sigma_{X}$) as high as the observations.

The exact number depends
on the exact choice of analysis method (see Sect.~\ref{robust}).
We find the best match with observations for mode lifetimes of
about 2 days, in good agreement with Paper I.
However, we note that there is also a reasonable match for all mode 
lifetimes less than a day, as their distributions
all look very similar to the bottom panel.
Randomly distributed peaks also show similar distributions to
$\tau=6\,$hours. Hence, if the mode lifetime of $\xi\,$Hya is only
a fraction of a day it would definitely destroy any prospects
for asteroseismology on this star.
We have confirmed this with simulations that do not include any comb
patterns by simulating a single mode with a mode lifetime of 2 hours,
similar noise, excess power hump, and with the same total power as
the above simulations.


\subsection{Robustness of method}\label{robust}

An important step for a correct interpretation of the results shown in
Fig.~\ref{fig3} is to test the robustness of our method.
We tested the dependence of the two relevant output parameters
$\Delta\nu_{0}$ and min($\sigma_{X}$) on the following:
\begin{enumerate}
 \item the initial guesses of $\Delta\nu_{0}$ and $X_{0}$
       in the minimization process of $\sigma_{X}$,
 \item the number of frequencies used to calculate $\sigma_{X}$,
 \item whether weighting of frequencies is used when calculating $\sigma_{X}$,
 \item the frequency separation of input frequencies,
 \item the intrinsic scatter of input frequencies,
 \item the method for measuring frequencies.
\end{enumerate}

We discuss each of these in turn in Appendix~\ref{robustapp},
but in summary, none of the listed points (1--6) have a significant
effect on the results shown in Fig.~\ref{fig3}.


\section{Power and frequency analysis}

\begin{figure}\centering
\includegraphics{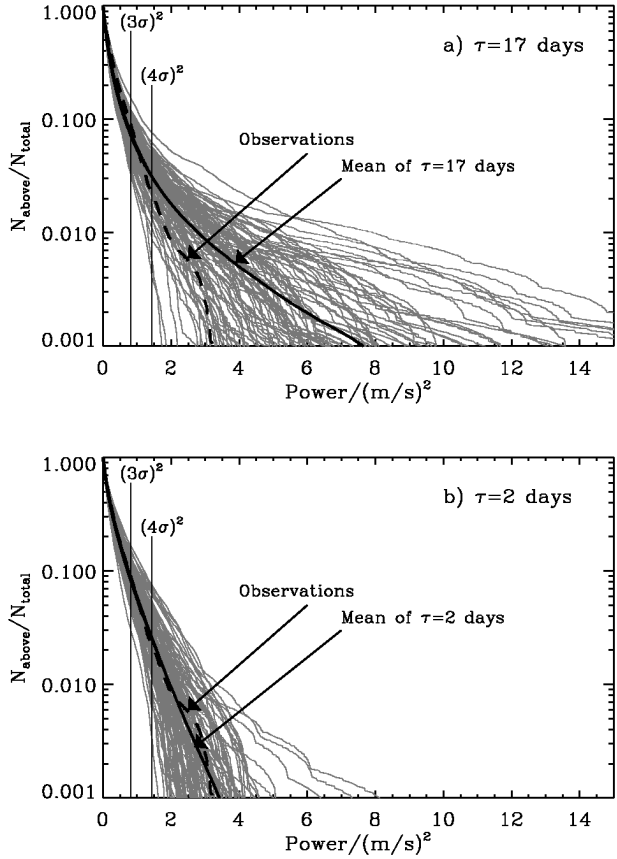}
\caption[Power distrib]{\label{fig6}
{\textbf{Panel (a):} Cumulative distribution of power spectra from 100 
simulations with
$\tau=17\,$days (grey lines). The average distribution is shown with
a black line. The dashed line indicates the distribution from the
observations.
\textbf{Panel (b):} Similar to panel (a), but with $\tau=2\,$days.
The noise level in the observed power spectrum is indicated.
}}
\end{figure}

To further investigate the mode lifetime for $\xi\,$Hya, we compared
the observed cumulative distribution of the power spectrum
from 1--190$\,\mu$Hz with that from simulations (Fig.~\ref{fig6}).
Note that the simulator is normalized so that on average 
it reproduces the observed power regardless of the mode lifetime 
(Paper I, Eq. 7).
For each level in power, we calculated the fraction of the power spectrum that
is above that level (similar to Fig. 2 in \citealt{DelacheScherrer83};
see also \citealt{Brown91,Bedding05,Bruntt05}).
We see only few time series of $\tau=17\,$days that have a power distribution
similar to that observed and on average the difference is significant,
while the distributions for $\tau=2\,$days resemble the observations much
better. We note that there is a large intrinsic variation seen
in the power distributions for $\tau=17\,$days.

\begin{figure}\centering
\includegraphics{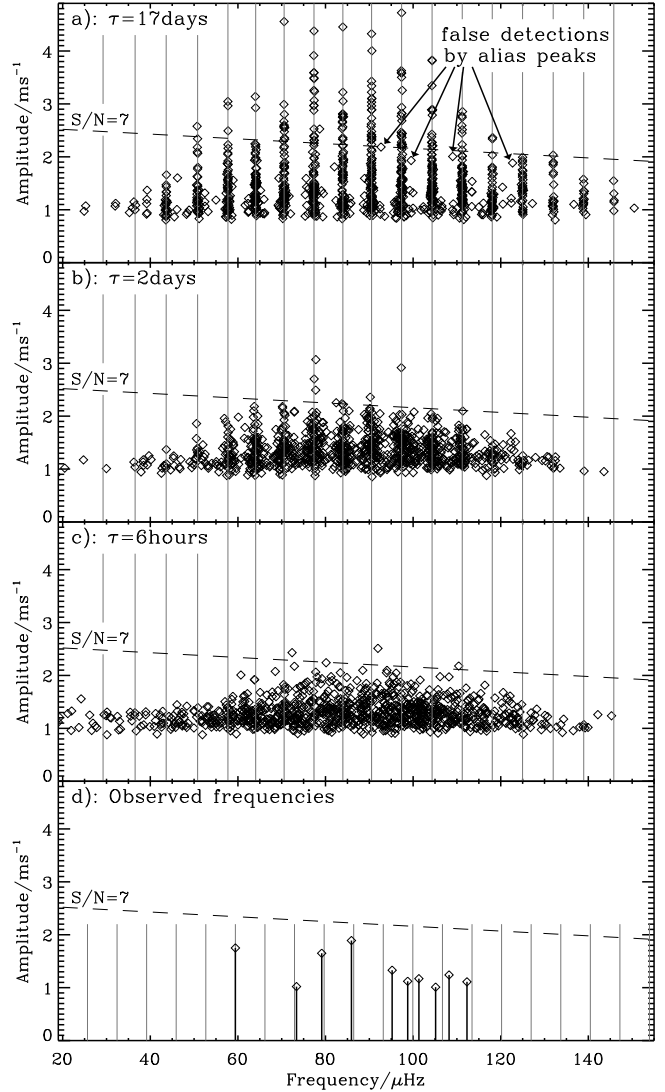}
\caption[unambiguity]{\label{fig7}
{Each panel shows 1000 measured frequencies and their amplitudes
(10 from each of 100 simulations). Dashed lines indicate 7 times the
input noise, and the solid grey lines are the input frequencies.
\textbf{Panel (a):} Mode lifetime = 17 days. 
\textbf{Panel (b):} Mode lifetime = 2 days.
\textbf{Panel (c):} Mode lifetime = 6 hours. 
\textbf{Panel (d):} Observed frequencies and the best matching comb pattern 
(grey lines).}}
\end{figure}

We also investigated the reliability of the observed frequencies
(Table~\ref{table1}), using the simulations
described in Sect.~\ref{sim}. The measured frequencies,
together with the input frequencies and noise level are plotted in Fig.~\ref{fig7}.
Apart from the broadening of the mode frequencies due to damping,
we also see false detections of alias peaks. This is most easily
seen in panel (a) where the damping is less (a few examples of false detections
are indicated).
Our test shows that frequencies are not unambiguous
if S/N$\lesssim$7--8, even for mode lifetimes of 17 days.
For short mode lifetimes it looks rather hopeless to measure
the individual mode frequencies with useful accuracy.
The observed frequencies are all with S/N$<$6.3 and hence cannot be
claimed to be unambiguous.
Hence the frequencies in Table~\ref{table1} (see also Fig.~\ref{fig7} 
bottom panel) should not be used for a direct compari\-son with
individual frequencies from a pulsation model.
The lack of a clear comb pattern similar to Fig.~\ref{fig7} (top panel)
in the observed frequencies also supports a short mode lifetime.


\section{Discussion and future prospects}\label{discussion}

Our best match for the mode lifetime of $\xi\,$Hya ($\tau\sim2\,$days)
is in good agreement with our estimate in Paper I but disagrees
with theory \citep{HoudekGough02}.
This discrepancy indicates that red giants could be used
to better understand the mechanisms of the driving and damping of
oscillations in a convective environment.
The short mode lifetime of $\xi\,$Hya also
suggests a narrow range of mode lifetimes
for a large range of stars, from the main sequence
to the red giants, and hence a steep decline in the `quality' factor
(Fig.~\ref{figteas}). The transition from the short mode lifetime
regime to the much longer lifetimes of the semi-regular variables
still needs to be investigated. It is likely
to involve some kind of interaction between pure stochastic excitation and
excitation by the $\kappa$ mechanism, which is responsible for
the oscillations we see in Mira stars \citep{Bedding03a}.

A difficulty in using the current observations of $\xi\,$Hya for 
asteroseismology
arises from the severe crowding in the power spectrum. We now discuss
possible origins of the crowding and some
aspects of this issue.

We do not expect crowding from high amplitude non-radial modes in 
the power spectrum although low amplitude modes cannot be excluded 
\citep[][ Paper I]{Dalsgaard04}. Additional simulations
are needed to quantify the effect on the lifetime estimate from low
amplitude non-radial modes, but we expect it to be small.

\begin{figure}\centering
\includegraphics{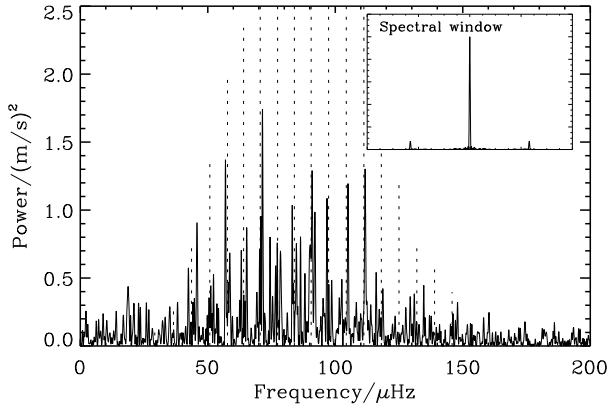}
\caption[sim multisite]{\label{fig8}
{Power spectrum of simulated time series of $\xi\,$Hya ($\tau=2\,$days)
using a window function from a two-site campaign where the sites are separated by
12 hours in longitude (see text). Dotted lines indicate the input frequencies.
}}
\end{figure}

The crowding in the power spectrum comes partly from the single-site 
spectral window,
which emphasizes the importance of using more continuous
data sets from multi-site campaigns or space missions.
To illustrate this, we construct the combined time series from a
hypothetical two-site
observing campaign on $\xi\,$Hya where the observing window from
each site is identical to that of the present data set obtained
with the \textsc{Coralie} spectrograph. The other site is assumed to be
a twin at the complementary longitude (12 hour time shift), and
the two individual data sets have identical sampling and noise.
We simulated the stellar signal using $\tau=2\,$days, with other
parameters the same as for Fig. 12 in Paper I.
A power spectrum of such a two-site time series is shown in
Fig.~\ref{fig8}.
Each mode profile is seen much more clearly, though slightly
blended due to the short mode lifetime.
Obviously, more can be obtained from such a spectrum than from
our present data set (Fig.~\ref{fig4}), but a thorough
analysis of similar simulations should be done to determine the prospects
for doing asteroseismology on $\xi\,$Hya.

As more high-quality asteroseismic data become available,
simulations will continue to be an important tool
for interpreting the data.
However, one has to be careful about what can be deduced
from simulations.
Even though we may think, from an ideal
perspective, that some parameters do not affect our measurements,
this may turn out not to be true.
A realistic noise source, the right window function, number-
and scatter of frequencies, their amplitude and mode lifetime
are all parameters that have to be considered
when simulating oscillation data.
Our comparison of the results from simplified simulations with more
comprehensive simulations shows clearly
that using simplified simulations can lead to erroneous conclusions.
As an example, we measured an increase by a factor of two of
min($\sigma_{X}$) between simulations including only 1 and 18 input frequencies.
This demonstrates that it is
important to use realistic (non-simplified) simulations to
relate frequency scatter to the mode lifetime.

A complete application of our method on other stars
(e.g. $\alpha\,$Cen A, $\nu\,$Ind, $\eta\,$Boo, and the Sun) will be
published in a forthcoming paper.


\section{Conclusions}\label{conclusion}

We introduced a new method that uses only the
scatter of the measured frequencies from a comb pattern to measure the mode
lifetime of the red giant $\xi\,$Hya.
This method takes into account false detections of
noise and alias peaks. 
We find that the most likely mode lifetime is about 2 days, and
we show that the theoretical prediction of 17 days \citep{HoudekGough02}
is unlikely to be the true value.

Due to the high level of crowding in
the power spectrum, the signature from the p-modes
is too weak to determine the
large separation to very high accuracy.
However, our measurement supports the separation
of $6.8\,\mu$Hz found by \citet{Frandsen02}.

We conclude that the only quantities we can reliably obtain from
the power spectrum of $\xi\,$Hya are the mode amplitude, mean mode
lifetime, and the average large frequency separation. 

Our simulations show that
none of the measured frequencies from
the $\xi\,$Hya data set \citep{Frandsen02} can be regarded as unambiguous.
Hence we suggest the measured frequencies given in Table~\ref{table1}
are not used for direct mode-matching with a pulsation model to
constrain the stellar model.
Only in the case of a greatly improved window function could
this be possible.

\begin{acknowledgements}
This work was supported in part by the Australian Research Council.
\end{acknowledgements}


\appendix

\section{CLEAN test}\label{clean}

The method described in Sect.~\ref{method} required extraction of many
frequencies. We therefore tested two different sine-wave fitting or
CLEANing methods,
simple CLEAN and CLEAN by simultaneous fitting. We call them CLEAN1
and CLEAN2, respectively. The tests were done on our simulated
time series to gain better understanding of the
results from CLEANing and to determine which method was most favorable
in our case. Both methods subtract
one frequency at a time (the one with highest amplitude), but
CLEAN2 recalculates the
parameters (amplitude, phases, and frequencies) of the
previously subtracted peaks while fixing the frequency of the
latest extracted peak. In this way, the fit of the sinusoids to
the time series is done simultaneously for all peaks.
CLEAN1, however, does not recalculate the 
parameters of previously subtracted peaks.
The time used by CLEAN2 is a factor of $1.5(N_{\mathrm{extract}}+1)$
longer than for CLEAN1, where $N_{\mathrm{extract}}$ is the number
of extracted frequencies.

We made simulations similar to those described in Sect.~\ref{sim} but
with 17 equally spaced frequencies and no noise added.
From each set of 100 time series with a given mode lifetime, 10 frequencies
were extracted providing 1000 frequencies which we
compared with the input frequencies.
For coherent oscillations CLEAN2 is doing a perfect job while
CLEAN1 detected 1\% false peaks. If only true detections are considered,
the scatter of the extracted frequencies relative to the input is
roughly equal to the frequency resolution ($1/T_{\mathrm{obs}}$) for
CLEAN2, while the CLEAN1 results scatter twice as much.

For non-coherent oscillations, the number of false detections
increases for
both methods and the relative success rates of the two methods
become equally good (or bad), for
mode lifetimes in the order of the observational window or shorter.
In this regime, the scatter of the extracted
frequencies gets dominated by the mode damping and is
independent of the CLEANing method.
For mode lifetimes below 5 days, the scatter of the frequencies
is so pronounced relative to the spacing between modes and the
alias peaks that they overlap and it cannot be
determined whether an extracted frequency is false or not
(see e.g. Fig.~\ref{fig1}b).

The amplitude of a detected frequency found by both CLEANing methods
generally differ by maximum +/-30\% but can for a few cases differ
by 70\%, being most severe for shorter mode lifetimes.
There is a general trend that CLEAN2 ascribes lower amplitudes
to the peaks of highest amplitude in a spectrum, and higher
amplitudes to low amplitude peaks relative to CLEAN1. This is a
result of the recalculation of the oscillation parameters in CLEAN2.
The detected amplitude difference was not considered as
crucial as we were interested in using the frequencies only.
Since there is observational evidence that the mode lifetime
of $\xi\,$Hya is below 5 days (Paper I), where the two methods 
perform equally well, we choose CLEAN1 due
to its faster algorithm (by a factor of 17 in our case where
$N_{\mathrm{extract}}=10$).


\section{Robustness tests}\label{robustapp}

In this appendix we discuss each of the points (1--6) listed in 
Sect.~\ref{robust} in turn:

(1) In the minimization process, the initial guesses of $\Delta\nu_{0}$
and $X_{0}$ could give quite different outputs for
$\Delta\nu_{0}$ on a data set with a shallow dip in $\sigma_{X}$
(i.e weak comb pattern).
For data with a deep dip in $\sigma_{X}$ (i.e. strong comb pattern),
$\Delta\nu_{0}$ is unaffected. We found that
min($\sigma_{X}$) was well determined if the initial guess
of $\Delta\nu_{0}$ was within $\sim 1\,\mu$Hz of
the true value.
As a representative example we see $\pm1.5\,\mu$Hz change in
$\Delta\nu_{0}$ and only $\pm0.008\,\mu$Hz change in min($\sigma_{X}$)
for the set of observed frequencies, which produced a shallow dip 
in min($\sigma_{X}$).
Our final setup (plotted in Fig.~\ref{fig3}) was to fix the initial
guess of $\Delta\nu_{0}$ at $6.8\,\mu$Hz, which we believe is within
$1\,\mu$Hz of the true value,
and to use 7 different $X_{0}$ values (separated $1\,\mu$Hz
apart). We then chose the solution with the lowest minimum found from
these 7 trials.

(2) We investigated the changes in the output parameters
(min($\sigma_{X}$), $\Delta\nu_{0}$), from varying the
number of measured frequencies from 5 to 10.
The average $\Delta\nu_{0}$ is unaffected by the number
of frequencies, but
more frequencies give a more accurate $\Delta\nu_{0}$
determination.
For the observed frequencies $\Delta\nu_{0}$ changes by
$\pm0.5\,\mu$Hz in the tested regime (from 5 to 10 frequencies).
Furthermore we observed higher values of min($\sigma_{X}$), thus
weaker comb patterns, with increasing number of frequencies, but
the relative difference between the min($\sigma_{X}$) distributions
from simulations and the observational data point did not change
significantly.

We decided to use 10 frequencies since we trusted our
min($\sigma_{X}$) distributions more when the $\Delta\nu_{0}$
determinations were closer to the true value. Measuring more than
10 frequencies showed no clear advantage, for this particular data set.

(3) Assigning weights to each frequency according to its S/N makes the absolute
value of min($\sigma_{X}$) slightly less sensitive to the number of
measured frequencies used to calculate min($\sigma_{X}$), which is expected.
Due to the higher weight given to the central part of the
damping profile for each mode \citep{Anderson90}, the frequency separation
is also slightly better determined using weights. However, min($\sigma_{X}$)
will scatter slightly more, making it more difficult to distinguish
min($\sigma_{X}$) distributions from different mode lifetimes.
This is probably a result of assigning lower weight to the tails of the
damping profile. We chose not to use weights, to obtain the best
possible result on the mode lifetime.

(4) Changing the large separation of the input frequencies did not produce any
change in min($\sigma_{X}$) for the tested range (6.8--7.2$\,\mu$Hz),
but only changed the output $\Delta\nu_{0}$ accordingly.
Hence, it is not crucial for our results on the mode lifetime
to know the true frequency separation to a very high accuracy.

(5) We also changed the deviation, $X(\nu_{i})$, from a comb pattern
of the input frequencies.
We used input frequencies that had a deviation twice as large
as in the pulsation model (see Fig.~\ref{fig0}),
and also a frequency set that followed a perfectly regular comb. No significant
change of the min($\sigma_{X}$) distributions (Fig.~\ref{fig3}) was
observed. We believe this is because min($\sigma_{X}$)
is dominated by damping for the mode lifetimes we investigated.

(6) Finally, we investigated whether the choice of
frequency extraction method affects min($\sigma_{X}$)
and $\Delta\nu_{0}$.
Using three different algorithms on the observations provided
3 different sets of measured
frequencies, and hence 3 different min($\sigma_{X}$) values that
ranged from 0.23--0.28. The corresponding output
$\Delta\nu_{0}$ ranged from 6.7--7.2$\,\mu$Hz.
We note, that the min($\sigma_{X}$) distributions seen in
Fig.~\ref{fig3} did not change significantly.


\bibliography{bib_complete}

\end{document}